\documentclass[aps,prl,twocolumn,showpacs,groupedaddress]{revtex4-1}

\usepackage{amsmath}
\usepackage{amssymb}
\usepackage{psfrag}
\usepackage{graphicx}

\newcommand {\e} {\text{e}}                  
\newcommand {\eps} {\varepsilon}                  
\newcommand {\dd} {d}                  
                   	
\newcommand {\w}{\omega}                       
                       
\newcommand{\inv}{^{-1}}
\newcommand{\eq}[1]{(\ref{eq:#1})}
\newcommand{\vv}[1]{\mathbf{#1}}

\newcommand{\OO}{\mathcal{O}}
\newcommand{\fig}[1]{\ref{fig:#1}}
\newcommand{\mean}[1]{\langle #1\rangle}

\newcommand{\del}[1]{ \partial_{#1}}
\newcommand{\ddd}[1]{\frac{d}{d#1}}
\newcommand{\intl}{ \int\limits}
\newcommand{\sqg}{ \sigma }
\newcommand{\Gam}{ \sigma^2 }

\newcommand{\ph}{\varphi}
\newcommand{\tht}{\theta}

\begin{document}

\title{Effective phase description of noise-perturbed and noise-induced oscillations}

\author{Justus T. C. Schwabedal}
\email{jschwabedal@googlemail.com}
\author{Arkady Pikovsky}
\affiliation{Department of Physics and Astronomy, Potsdam University, 14476 Potsdam, Germany}

\begin{abstract}
An effective description of a general class of stochastic phase oscillators is
presented. For this, the effective phase velocity is defined either by invariant
probability density or via first passage times. While the first approach
exhibits correct frequency and distribution density, the second one yields
proper phase resetting curves. Their discrepancy is most pronounced for
noise-induced oscillations and is related to non-monotonicity of the phase
fluctuations.
\end{abstract}

\maketitle

\section{General introduction} 

The phase is a theoretical concept lying at the heart of description of
oscillatory dynamics. In a physical context a periodic phase variable provides
a simplified description of self-sustained oscillations shown by natural,
synthetic or mathematical systems in their state space
\cite{Kuramoto-84,Pikovsky-Rosenblum-Kurths-01}. The phase description contains
many characterizing physical properties associated to oscillatory motion as for
example frequency and regularity of oscillations. Most importantly, smooth or
impulsive coupling of interacting oscillators may be formulated in terms of
phase dynamics~\cite{Kuramoto-84,Kralemann-08,Canavier2007}.

Irregular features, interpreted as noise, may be present in oscillations. In
many situations noise can be treated as a perturbation to oscillatory dynamics.
In this case one can start with the phase description of noiseless oscillations,
and consider noise as a perturbation. However, noise may be  substantial in the
sense that it induces oscillations in the system which would equilibrate
otherwise. Such noise-induced oscillations may be quite coherent disguising the
underlying systems excitable nature. Therefore, it may not be possible to
distinguish between the two cases in an experimental setup unless noise may be
eliminated.

Despite of evident similarities between noise-perturbed and noise-induced
oscillations, a phase description of noise-induced oscillations cannot be
obtained through standard perturbative procedures, because in the noise-free
situation there is no dynamics. Recently, this problem was addressed by an
effective description, where noise-induced oscillations driven by an external
periodic force could be characterized in a genuine way~\cite{Pikovsky2010}. It
was seen, that the effective phase model relies on an average concept of speed
given by the current velocity~\cite{Nelson-66}.

The goal of this paper is to generalize the method of~\cite{Pikovsky2010}. For
this, we show that an alternative effective phase model is possible, and
compare the two approaches. After a summary of well-known dynamical properties
of stochastic phase oscillators in the next section, we outline the current
model of effective phase theory for a single oscillator (Section
\ref{sec:currentModel}). Hereon, an alternative model based on first passage
times is presented in Section \ref{sec:ptm} modeling other aspects of
stochastic dynamics. In Section \ref{sec:asymptotics}, mathematical aspects of
the theory in its limiting cases for small and large noise are discussed
highlighting differences and similarities of the effective phase models. A
special emphasis is put on the description of noise-induced oscillations in
these limits. In Section \ref{sec:stochPRC}, phase resetting curves of
stochastic phase oscillators and their relation to the first passage model of
effective phase theory are described.

\section{Stochastic phase oscillators and their effective description} 

Our basic model is a stochastic phase oscillator. It is described by a
$2\pi$-periodic random process $\theta$, called \textit{protophase}, that
obeys the Langevin dynamics
\begin{equation}
	\dot{\tht}(t)=h(\tht(t))+g(\tht(t))\xi(t),
	\label{eq:generalOsci}
\end{equation}
where $\xi(t)$ is $\delta$-correlated Gaussian noise $\langle \xi(t)\xi(t')\rangle=2\delta(t-t')$. A well-known example
showing most prominent features of stochastic oscillations is the theta model~\cite{Ermentrout2008}
\begin{equation}
	\dot{\tht}(t)=a+\cos\tht+\sqg\xi(t)~.
	\label{eq:thetaNeuron}
\end{equation}
For $|a|<1$, it shows noise-induced oscillations which remain also in the deterministic limit $\Gam\to0$, while for $|a|>1$
oscillations are noise-perturbed and they persist also for vanishing noise.

\begin{figure}[h]
   \centering
   \includegraphics[width=0.49\textwidth]{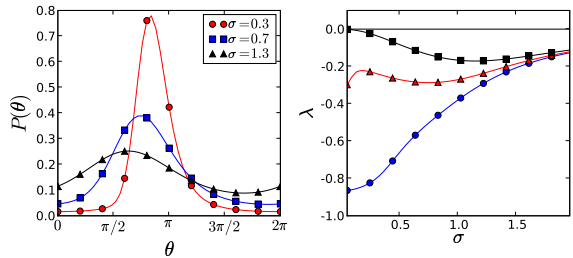}
   \caption{ Left: Probability density \eq{FPQ-sol} of the theta-model for $a=0.9$ and 
   		values of $\sqg$ as indicated. Right: Lyapunov exponent \eq{lyap} of theta-model as
		a function of $\sqg$ for $a=0.5$ (blue circles), $a=0.95$ (red triangles), and $a=1.5$ (black squares). }
   \label{fig:theta-Pl}
\end{figure}

For the stochastic phase oscillator introduced above, 
most relevant quantities are accessible
analytically. The probability density $P(\tht)$ is governed by the
Fokker-Planck equation associated to equation \eq{generalOsci} given by
\begin{equation}
	\partial_tP=-\del{\tht}\left[hP\right]+\del{\tht}\left[ g\del{\tht}\left[gP\right]\right]=-\del{\tht}J~.
	\label{eq:FPQ}
\end{equation}
For stationary probability density, the \textit{probability flux} $J$ is
constant and we obtain the simpler equation
\begin{equation}
	J=hP-g\partial_\tht\left[gP\right]~.
	\label{eq:flux}
\end{equation}
This equation has the well-known solution~\cite{Risken1989}
\begin{equation}
	P(\tht)=C\int_\tht^{2\pi+\tht}\frac{\dd\psi}{g(\tht)g(\psi)}~\e^{-\int_\tht^\psi\frac{h(\ph)}{g^2(\ph)}~\dd\ph}~,
	\label{eq:FPQ-sol}
\end{equation}
where $C$ is a \textit{normalization constant} ensuring
$\int_0^{2\pi}P(\tht)~\dd\tht=1$. As illustrated in Fig.~\fig{theta-Pl},
probability density becomes singular for vanishing noise if oscillations are
noise-induced.

\begin{figure}[h]
   \centering
   \includegraphics[width=0.49\textwidth]{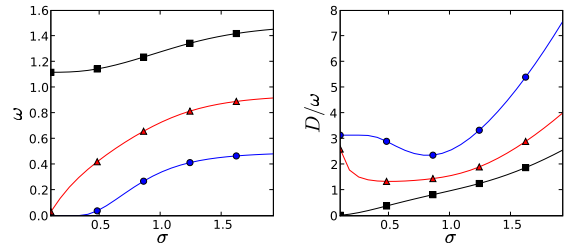}
   \caption{ Mean frequency \eq{freqflux} in left plot, and quotient of diffusion coefficient \eq{diffCoef} 
   	and mean frequency in right plot of the $\theta$-model as functions of $\sigma$ for $a=0.5$ (blue circles),
	$a=0.95$ (red triangles), and $a=1.5$ (black squares). }
   \label{fig:theta-wD}
\end{figure}

Among the traditional quantities of stochastic phase oscillators are the mean
frequency and diffusion coefficient
\begin{equation}
	\w=\lim_{t\to\infty}\frac{\mean{\theta(t)}}{t}~,~\text{and}~~D=\lim_{t\to\infty}\frac{\mean{\left[\tht(t)-\w t \right]^2}}{2t}~,
	\label{eq:freqDiff}
\end{equation}
which actually are properties of the point process $n(t)$ counting the number
of rotations of $\tht(t)$. They are expressed in terms of $h(\tht)$ and
$g(\tht)$ through the well-known formulas~\cite{Risken1989,Reimann2001}
\begin{equation}
	\w = 2\pi J = 2\pi C\left[1-\e^{-\int_0^{2\pi}\frac{h(\ph)}{g^2(\ph)}~\dd\ph}\right]~\text{, and}
	\label{eq:freqflux}
\end{equation}
\begin{equation}
	D=\frac{\frac{1}{2\pi}\int_0^{2\pi}\frac{d\psi}{g(\psi)}
	\left[\int_{\psi-2\pi}^\psi\frac{d\ph}{g(\ph)}\rho(\ph,\psi)\right]^2
	\int_\psi^{\psi+2\pi}\frac{d\ph}{g(\ph)}\rho(\psi,\ph)}
	{\left[\frac{1}{2\pi}\int_0^{2\pi}\frac{d\psi}{g(\psi)}\int_{\psi-2\pi}^\psi\frac{2d\ph}{g(\ph)}\rho(\ph,\psi)\right]^3}~,
	\label{eq:diffCoef}
\end{equation}
where $\rho(\theta,\ph)=\exp\left[-\int_{\theta}^{\ph}
\frac{h(\eta)}{g^2(\eta)}~\dd\eta\right]$. For noise-induced oscillations the
mean frequency converges to zero in the limit of vanishing noise, as shown in
the left plot of Fig.~\fig{theta-wD}. In the large noise limit
$\Gam\to\infty$ the mean frequency converges to a finite value for additive
noise, which will be shown in a general setup in section \ref{sec:asymptotics}.
The quotient of diffusion coefficient and mean frequency is a measure for
decoherence of oscillations (right plot). For noise-induced oscillations, the
well-known effect of coherence resonance is observable where decoherence
decreases for increasing noise amplitude (red triangles and blue
circles)~\cite{Pikovsky-Kurths-97}. The effect is not observable for
noise-perturbed oscillations (black squares).

Another interesting quantity that characterizes phase oscillations is Lyapunov
exponent $\lambda$ associated to noise. It quantifies whether oscillators under
the influence of the same noise representation $\xi(t)$ will synchronize by
stochasticity. For oscillator \eq{generalOsci} it is computed by
\begin{equation}
	\lambda = \mean{h'(\tht)+g''(\tht)g(\tht)}.
	\label{eq:lyap}
\end{equation}
For oscillators under the influence by additive noise like the theta model,
$\lambda$ vanishes in the limit $\Gam\to\infty$. For small noise there are
three cases as exemplified in figure (\ref{fig:theta-wD}). For a
noise-perturbed oscillator (black squares) the Lyapunov exponent goes to zero
as $\lambda\propto-\Gamma$ \cite{Goldobin2005}. For excitable oscillators
(red triangles and blue circles) it converges to a finite negative value
dependent on the quantitative stability of their fixed point. For these, there
arises yet the special case where $\lambda$ shows a local minimum as a function
of $\sqg$ (red triangles) which somehow resembles the noise-perturbed case.


As we have seen, phase dynamics of noise-induced oscillators can be quite close
to that of periodic ones, what suggests a purely deterministic phase equation of
the type
\begin{equation}
	\dot{\tht}=F(\tht)~,
	\label{eq:genefd}
\end{equation}
for a theoretical description to be possible. Generally, one cannot set $F(\theta)=h(\theta)$
by taking the deterministic part of the Langevin model~(\ref{eq:generalOsci}),
as $h(\theta)$ may have zeros and thus $\theta$ would be non-oscillating.
Instead, we have to construct an \textit{effective phase model} using some
criteria to determine $F$.  Generally, we demand that the effective phase model
of type (\ref{eq:genefd}) represent as many characteristic properties of
stochastic phase oscillators as possible.  In~\cite{Pikovsky2010}, one approach
was proposed that allowed for a construction of an effective phase model with
the same mean frequency (i) (equivalently, the same mean period) and
distribution density (ii) as the stochastic phase oscillator. Without violating
these conditions, the diffusion coefficient (iii) could be modeled, too, by
adding noise to (\ref{eq:genefd}) in a certain way. Drawing a more general
framework, we will present in the next two sections the effective phase model
from \cite{Pikovsky2010} and another effective phase model based on first
passage times in a coherent way.  Roughly, the difference of these models is in
the definitions of the mean velocity. Given an interval of length $\Delta\tht$,
velocity can be measured as the quotient of $\Delta\tht$ and the mean time
$\Delta t$, that $\tht(t)$ spends in this interval. This leads exactly to an
effective phase description as presented in~\cite{Pikovsky2010}. Alternatively,
velocity can be measured as the quotient of $\Delta\tht$ and the mean time
taken to reach the opposite boundary of the interval, leading to the concept of
\textit{first passage velocity}. For deterministic phase oscillators the two
definitions of velocity coincide, whereas for oscillations driven by a random
force there is a difference, which is especially pronounced for oscillations
that are noise-induced.

\section{Current Model of effective phase dynamics} \label{sec:currentModel}

\begin{figure}[h]
   \centering
   \includegraphics[width=0.49\textwidth]{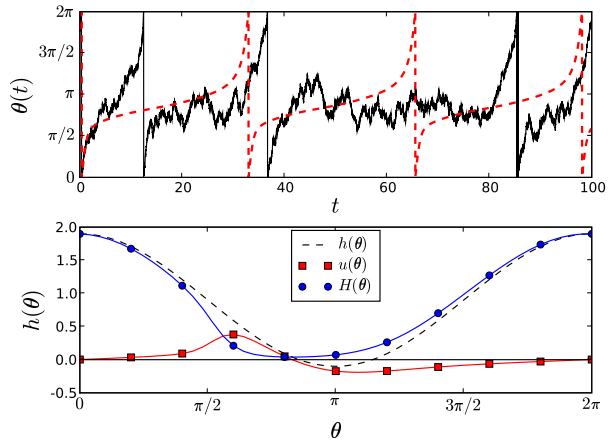}
   \caption{ Current model \eq{effOsci} has dynamics $\tht(t)$ strongly resembling
   	that of the theta model \eq{thetaNeuron}, here at $a=0.9$ and $\sqg=0.3$ (top).
   	For noise-induced oscillations the osmotic contribution $u(\tht)$ to current velocity~\eq{Heff2}
	is non-negligible and accounts for its strict positivity. }
   \label{fig:sec3-flowModel}
\end{figure}

We start by constructing the deterministic phase equation for the
\textit{current model}
\begin{equation}
	\dot{\tht}=H(\tht)~,
	\label{eq:effOsci}
\end{equation}
by using the notion of speed based on the mean time that the stochastic phase
oscillator \eq{generalOsci} spends in an interval $[\tht,\tht+\dd\tht]$. This
time is defined according to the invariant probability density $\dd
t=P\dd\tht/J$. It follows, that the \textit{current velocity} $H$ is given by
\begin{equation}
	H(\tht) =\frac{J\dd\tht}{P(\tht)\dd\tht} = \frac{\w}{2\pi P(\tht)}~.
	\label{eq:flowvelocity}
\end{equation}
This leads to the same effective model introduced in \cite{Pikovsky2010}.
Model \eq{effOsci} obeys conditions
(i)[preservation of the mean frequency] and (ii) [preservation of the probability density]
because it fulfills the stationary Liouville equation for $P(\tht)$
and it shows correct period $T$ as seen by
\begin{equation}
	-\del{\tht}\left[ H(\tht)P(\tht)\right]=0~,~\text{and}~\int_0^{2\pi}\frac{d\tht}{H(\tht)}=T~.
	\nonumber
\end{equation}
In Fig.~\fig{sec3-flowModel} we compare time series of oscillator \eq{thetaNeuron} and its current model
\eq{effOsci}. Although the stochastic components are
missing in the effective model, the observed dynamics is comparable.

The current velocity $H(\theta)$ can be expressed in terms of $h(\tht)$ and $g(\tht)$. For
this, equation \eq{flux} is divided by $P$, and the result is compared to
equation \eq{flowvelocity} yielding
\begin{equation}
	H(\tht) = h(\tht)-\frac{1}{2}\left[g^2(\tht)\right]'-g^2(\tht)\left[\ln P(\tht)\right]'=h(\tht)-u(\tht)~.
	\label{eq:Heff2}
\end{equation}
It consists of the deterministic contribution $h(\tht)$ and an osmotic
contribution $u(\tht)$, that is especially pronounced for the noise-induced
oscillations in the theta model (Fig.~\fig{sec3-flowModel}
bottom panel). The current velocity corresponds to the point-wise average of
central differences~\cite{Just2003}. Therefore, it may be constructed from an observed (e.g., experimentally)
time series $\tht_n=\tht(n\Delta t)$ by a simple averaging procedure
\begin{equation}
	H(\tht)\approx\left.\frac{\langle\tht_{n+1}-\tht_{n-1}\rangle}{2\Delta t}\right|_{\tht_n=\tht}~.
	\label{eq:Heff-esti}
\end{equation}

Having constructed the current model, we can transform the protophase $\theta$ to a uniformly rotating phase variable $\ph$, that describes oscillations in an invariant way. It has
simple properties
\begin{equation}
	\dot{\ph}=\w~;~~~P(\ph)=\frac{1}{2\pi}~.
	\label{eq:uniformPhase}
\end{equation}
As it can be easily checked, the nonlinear transformation $\tht\to\ph$ is given by
\begin{equation}
	\ph=S(\tht)=2\pi\int_0^{\tht}P(\tht')~d\tht'~.
	\label{eq:trafo1}
\end{equation}
With the transformation, coordinate dependent differences in the protophase $\Delta\tht=\tht_2-\tht_1$
can be transformed to invariant differences in the phase given by
\begin{equation}
	\Delta\ph=S(\tht_2)-S(\tht_1)=2\pi\int_{\tht_1}^{\tht_2}P(\tht)~\dd\tht~. 
	\label{eq:trafo1-2}
\end{equation}
Given a set of data $\tht_n=\tht(n\Delta t)$ containing
$N$ data points, transformation can be obtained numerically as
in~\cite{Kralemann-08}. If one is not interested in the transformation
$S(\tht)$ but in the transformed data $\ph_n=S(\tht_n)$ only, one may alternatively
evaluate
\begin{equation}
	\ph_n=\frac{2\pi}{N}\sum_{l=0}^{N-1}\Theta(\tht_n-\tht_l)~,
	\label{eq:trafo1-esti}
\end{equation}
which is implemented quickly by sorting ($\Theta$ is the Heaviside function).

It was also shown in~\cite{Pikovsky2010}, that simultaneously to the mean period and the probability density, phase diffusion may be
modeled. For this, $\delta$-correlated noise $\eta(t)$ is
added to the invariant phase dynamics \eq{uniformPhase} with noise intensity
$\sqrt{D}$ leading to
\begin{equation}
	\dot{\ph} =\w+\sqrt{D}\eta(t)~.
	\label{eq:effNoise}
\end{equation}
Now, $\ph$ has a diffusion constant $D$ while preserving uniform density and mean frequency $\w$. Therefore, application of the inverse
transformation $\tht=S\inv(\ph)$ gives us the stochastic current model
\begin{equation}
	\dot{\tht}=H(\tht)+\frac{\sqrt{D}}{\w}H(\tht)\eta(t)~,
	\label{eq:stochProtoPhase}
\end{equation}
that fulfills conditions (i) and (ii) for any value of $D$. It may be chosen
freely, and we chose it uniquely from condition (iii): Diffusion coefficients
of stochastic current model \eq{stochProtoPhase} and stochastic phase
oscillator \eq{generalOsci}, should be equal. This condition is fulfilled if
diffusion coefficient \eq{diffCoef} is used for $D$.


\section{Passage time model of effective phase dynamics}
\label{sec:ptm}

As an alternative to the current model outlined in last section, a velocity based on
first passage time statistics of oscillator \eq{generalOsci} shall lead us to a
\textit{first passage model}
\begin{equation}
	\dot{\tht} = N(\tht)~.
	\label{eq:fpm}
\end{equation}
To determine $\dot{\tht}\approx \frac{\Delta\tht}{\Delta t}$ we interpret $\Delta t$ as the first passage time of passing interval $\Delta \tht$. More precisely,  the \textit{first passage velocity} is constructed using the mean first passage
time $T(\alpha,\beta)$ which it takes for $\tht(t)$ to reach a boundary $\beta>\alpha$
starting at $\alpha$. We get
\begin{equation}	\frac{1}{N(\tht)}=\frac{dt}{d\tht}=\lim_{\eps\to0}\frac{T(\tht,\tht+\eps)-T(\tht,\tht)}{\eps}=\del{\beta}T(\tht,\tht)~.
	\label{eq:fpv}
\end{equation}
It fulfills condition that the mean frequency of (\ref{eq:fpm}) should be equal to the mean frequency of the oscillations, because the mean period $T$ is nothing else as $T(\tht,2\pi+\tht)$ and therefore $\int_0^{2\pi}d\tht/N(\tht)=T$. However, the condition
(ii) above [density in the effective model is equal to the density of stochastic oscillations] is generally not fulfilled, as will be shown next.

The distribution density of model \eq{fpm}, which we will call \textit{first
passage density}, is given by
\begin{equation}
	R(\tht)=\frac{\w}{2\pi N(\tht)}~.
	\label{eq:fpd}
\end{equation}
In order to derive an equation for $R(\tht)$, an equation for the mean first
passage time has to be established~\cite{Risken1989}. Consider Fokker-Planck equation \eq{FPQ},
with the sharp initial condition $P(\theta,0)=\delta(\theta-\alpha)$. In this
case equation \eq{FPQ} describes the conditional probability
$P(\theta,t|\alpha,0)$. The boundary conditions
\begin{equation}
	P(-\infty,t|\alpha,0)=P(\beta,t|\alpha,0)=0~,
	\label{eq:bCond}
\end{equation}
are introduced, which correspond to the fact that trajectories starting at $\theta(0)=\alpha$ should only be considered as long as
they do not reach boundary $\beta$. Now, $P$ has to be reinterpreted since the
normalization condition does not hold anymore. The \textit{no-passage
probability} $G(\alpha,t)$ is defined as the probability that at time $t$
boundary $\beta$ is not reached when starting at $\alpha$. It is defined as
\begin{equation}
	G(\alpha,t) = \int_{-\infty}^{\beta}P(\theta',t|\alpha,0)\dd\theta'~.
	\label{eq:chap2-8}
\end{equation}
By a backward-Kolmogorov expansion of the Markovian transition probability $P$ it is derived~\cite{Risken1989} that $G$ obeys 
\begin{equation}
	\del{t}G=h(\alpha)\del{\alpha} G+g(\alpha)\del{\alpha}\left[g(\alpha)\del{\alpha} G\right]~.
	\label{eq:bkexp}
\end{equation}
For $t\in[0,\infty)$ and $\alpha<\beta$, a distribution density
for random first passage times is given by $g(\alpha,t)=-\del{t}G$. With
respect to it, the mean first passage time
is given by
\begin{equation}
	T(\alpha,\beta) = \langle t\rangle = \int_0^{\infty}G(\alpha,t')~\dd t'~.
	\label{eq:chap2-10}
\end{equation}
Integrating equation \eq{bkexp} over positive times one obtains 
an equation for the mean first passage time
\begin{equation}
	\begin{aligned}
		-1 &= h\del{\alpha}T+g\del{\alpha}\left[g\del{\alpha}T\right]~.
	\end{aligned}
	\label{eq:meanT}
\end{equation}
Because $\del{\beta}T(\tht,\tht)=-\del{\alpha}T(\tht,\tht)=1/N(\tht)$, equation
\eq{meanT} may be rewritten for the first passage density $R(\tht)$ as
\begin{equation}
		J = hR+g\del{\tht}\left[gR\right]~,
\label{eq:densR}
\end{equation}		
solved by
\begin{equation}
		R(\tht) = C\intl_{\tht-2\pi}^{\tht}\frac{\dd\psi}{g(\tht)g(\psi)}\e^{-\int_{\psi}^{\tht}\frac{h(\ph)}{g^2(\ph)}\dd\ph}~,
	\label{eq:fpdeq}
\end{equation}
with a normalization constant $C$, and $J=\w/2\pi$. Note that, equation
\eq{densR} is similar to \eq{flux}, but has an opposite sign of the second term. It provides an easy to handle analytic
formula for the first passage velocity $N(\tht)$.

\begin{figure}[h]
   \centering
   \includegraphics[width=0.49\textwidth]{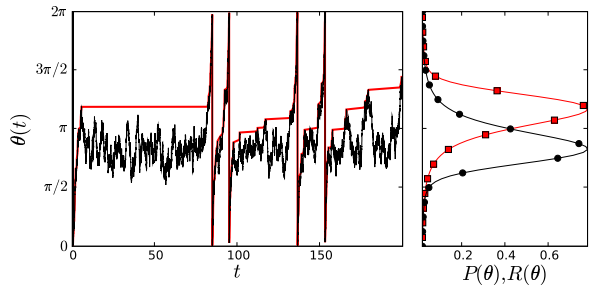}
   \caption{ Protophase $\tht(t)$ (black line) of theta model \eq{thetaNeuron} and its envelope (red thick line)
   		have similar long term dynamics (left plot), but their distribution densities
		$P(\tht)$ (black circles) and $R(\tht)$ (red squares) are different,
		especially for noise-induced oscillations, shown at parameters $a=0.9$ and $\sqg=0.3$. }
   \label{fig:sec3-fpd}
\end{figure}

The passage time density has a direct meaning for the stochastic protophase
$\tht(t)$, as we will explain in the following. Let $t_n$ be the times of first
passage, for which $\tht(t<t_n)<\tht(t_n)$ holds. This is illustrated in left
panel of Fig.~\fig{sec3-fpd} where realization $\tht(t)$ and point process
$\tht_n=\tht(t_n)$ are counterposed. Because of the Markov property, $\tht_n$
gives the starting point for a measurement of passage time ending when
$\tht(t)$ reaches $\tht_{n+1}=\tht_n+\dd\tht$. Although the trajectory of
corresponding time segment $\tht(t_n<t<t_{n+1})$ lies in the whole region
$\tht<\tht_{n+1}$, for the  first passage
density $R(\tht)$ it is attributed to the interval $\tht_{n}<\tht<\tht_{n+1}$.
So, in fact $R(\tht)$ is the density of the \textit{envelope} of the stochastic protophase $\tht(t)$, shown as a red bold line in Fig.~\ref{fig:sec3-fpd}.
 As shown in the right panel, the
probability density $P(\tht)$ and the first passage density $R(\tht)$ can be quite
different. The fact that $R(\tht)$ is probability density of the envelope of
$\tht(t)$ can be used for a numerical reconstruction from data, as it is shown
later in this section.

We would like to mention here that going from the stochastic protophase
$\tht(t)$ to its envelope, we achieve a monotonically growing protophase.
Indeed, phase is often understood as a strictly monotonic variable, in some
sense a "replacement" for a time variable. For a stochastic oscillator one
often observes "reverse" variations. Thus, taking the envelope is a natural way
to restore a monotonic  function of time.\\

\begin{figure}[h]
   \centering
   \includegraphics[width=0.49\textwidth]{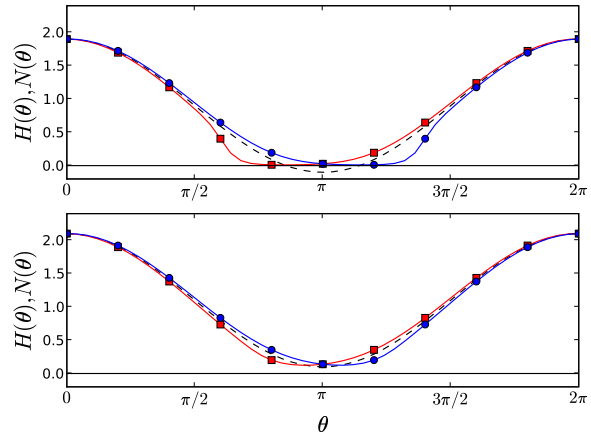}
   \caption{ Current velocity \eq{flowvelocity} (red squares) and first passage velocity \eq{fpv} (blue circles) for the theta model \eq{thetaNeuron} differ essentially,
   		for both noise-induced at $a=0.9$ (top plot), and noise-perturbed at $a=1.1$ (bottom plot) oscillations. Noise amplitude is $\sqg=0.2$.
		The difference in effective velocities is especially pronounced for noise-induced oscillations and does not disappear for small $\Gam$. }
   \label{fig:sec3-cmpRN}
\end{figure}

The first passage model \eq{fpm} provides effective phase dynamics alternative
to the current model. It fulfills condition (i) in that it shows the same mean
frequency as oscillator \eq{generalOsci}, but instead of modeling probability
density (ii), preserves first passage density (iib), which is for
deterministic oscillators equal to distribution density.  In Fig.~\fig{sec3-cmpRN}, first passage (blue circles) and current velocity (red
squares) are counterposed for theta models \eq{thetaNeuron}, for cases of
noise-induced (top panel) and noise-perturbed oscillations (bottom panel). In the
latter $N(\tht)$ and $H(\tht)$ coincide for vanishing noise, whereas for
noise-induced oscillations the difference widens.\\\\

In the numerical example, current and first passage velocities are mapped to each
other by mirror symmetry. This is due to the fact that both $h$ and $g$ are
symmetric in the theta model. For a stochastic phase oscillator with with symmetric $h$
and $g$, the transformation $\tht\to-\tht$ transforms equation \eq{densR} in
\eq{flux}. Their solutions are mapped to each other too. Therefore, symmetry
of $h$ and $g$ implies $N(\tht)=H(-\tht)$, as observed in Fig.~\fig{sec3-cmpRN}.

For the first passage model, a uniformly rotating phase variable is constructed
by the transformation
\begin{equation}
	\psi = Z(\tht)=2\pi\intl_0^{\tht}R(\tht')~\dd\tht'~.
	\label{eq:trafo2}
\end{equation}
Again, differences in protophase $\tht_2-\tht_1$ can are transformed to differences
in phase by
\begin{equation}
	\Delta\psi=Z(\tht_2)-Z(\tht_1)=2\pi\int_{\tht_1}^{\tht_2}R(\tht)~\dd\tht~. 
	\label{eq:trafo2-2}
\end{equation}

As for the current model, noise may then be taken into account by
\begin{equation}
	\dot{\tht}=N(\tht)+\frac{\sqrt{D}}{\w}N(\tht)\eta(t)~.
	\label{eq:Sfpm}
\end{equation}
This stochastic first passage model fulfills conditions (i), (iib),
and (iii). However, it does not fulfill condition (ii), as its probability
density is $R(\tht)\neq P(\tht)$. Therefore, an application of transformation
\eq{trafo2} to the stochastic variable $\tht(t)$ does not lead to a uniformly
distributed phase $\psi(t)$ as in equation \eq{uniformPhase}. 

The fact, that $R(\tht)$ is the probability density of the envelope of a
realization $\tht(t)$, can be used to obtain transformation $Z(\tht)$ from a
dataset $\tht(n\Delta t)$. The envelope of data has to be constructed. For
this, we need to find the times of first passages $t_n$ for which data has a
history that is strictly smaller, i.~e. $\tht(j\Delta t<t_n)<\tht(t_n)$. At
these first passages $\tht_n=\tht(t_n)$, transformation \eq{trafo2} is
estimated by formula
\begin{equation}
	\psi_n=Z(\tht_n) = \frac{2\pi}{T}\sum_{\tht_j<\tht_n}t_j-t_{j-1}~.
	\label{eq:trafo2-esti}
\end{equation}
To find the transformation on the whole domain one can proceed by an
appropriate interpolation of $Z(\tht_n)$ as for example smoothing splines.
Note, that equation \eq{trafo2-esti} gives a biased estimator. This can
be fixed by inserting a central difference scheme $(t_{j+1}-t_{j-1})/2$.


\section{Asymptotic properties} \label{sec:asymptotics}

\subsection{Singular perturbation for small noise} \label{ssec:singPerturb}

In the Fokker-Planck equation \eq{FPQ}, noise amplitude $g(\tht)$ appears in
front of the derivative with highest order in $\tht$. Therefore, a perturbation
expansion in $g(\tht)$ for a deterministic approximation of dynamics \eq{generalOsci}
might be singular. For an oscillator with additive noise 
\begin{equation}
	\dot{\tht} = h(\tht)+\sqg\xi(t)~,
	\label{eq:addNoise}
\end{equation}
we show that perturbation expansion becomes singular if oscillations are
noise-induced. We discuss the case in which the approximation of dynamics \eq{generalOsci}
should model the distribution density of oscillator
\eq{addNoise}. The singularity that arises in the limit $\sigma\to0$ is
discussed by perturbation expansion starting from $\sigma=0$, and starting from
$\sigma>0$ by effective phase theory. For the latter, current model \eq{effOsci}
is used.

For $\sigma=0$, our considerations start with the arbitrary model
$\dot{\theta}=h(\theta)$ for oscillator \eq{addNoise}. Suppose, it shows
noise-perturbed oscillations, i.~e. $h(\tht)$ strictly positive. Then, the
distribution density of arbitrary model gives a proper zeroth order
approximation to the probability density
\begin{equation}
	P(\theta)=\frac{C}{h(\theta)}+\OO(\sqg)~.
	\nonumber
\end{equation}
Better approximations can be obtained analytically by Taylor expansion of
$P(\tht)$. On the other hand, oscillator \eq{addNoise} may show noise-induced
oscillations, for which we impose without loss of generality that $h(\tht)$
shows two zero crossings. Then, the model has a stable fixed point at
$\tht_-$, and an unstable at $\tht_+>\tht_-$, and its distribution density is
given by
\begin{equation}
	P(\theta)=\delta(\tht-\tht_-)+\OO(\sqg)~,
	\label{eq:addNoise-Pex}
\end{equation}
corresponds to the probability density of equation \eq{addNoise}. However,
higher order terms in $\sigma$ that should lead to a smooth distribution
density must necessarily be singular. It is seen, that perturbation theory
becomes singular for noise-induced oscillations. Note, that first passage
density also shows a singular limit for noise-induced oscillations, however,
not at the stable, but the unstable fixed point. In the above example
it is given by
\begin{equation}
	R(\tht)=\delta(\tht-\tht_+)+\OO(\sqg)~.
	\label{eq:addNoise-Rex}
\end{equation}
On the other hand, first passage density and probability density converge for
noise-perturbed oscillations, reflecting the fact that for deterministic
self-sustained oscillators the quantities are synonym.

Starting from a current model to oscillator \eq{addNoise} computed at finite
$\sigma$, another view can be gained on the limit $\sigma\to0$. With formula
\eq{freqflux} and \eq{flowvelocity}, current velocity corresponding to \eq{addNoise} is
given by
\begin{equation}
	H(\tht) = \frac{1-\e^{-\frac{r(0,2\pi)}{\Gam}}}{\int_\tht^{\tht+2\pi}\frac{\dd\psi}{\Gam}\e^{-\frac{r(\theta,\psi)}{\Gam}}}~,~\text{with}~r(\theta,\psi) = \int_\tht^\psi h(\ph)~\dd\ph~.
	\label{eq:Heff-addNoise}
\end{equation}
Let us assume without loss of generality that $r(0,2\pi)$ is positive. In the limit of small
noise, the integral in the denominator of equation \eq{Heff-addNoise} is
dominated by the minimum of $r(\theta,\psi)$ with respect to $\psi$. Using the
method of stationary phase the zeroth order expansion
\begin{equation}
	H\left( \theta \right) = \OO(\sigma)+\left\{ 
		\begin{gathered}
			h(\tht)~:\text{ if } \text{argmin}_\psi \left[r(\tht,\psi)\right]=\tht\hfill \\
			0 ~ ~ ~ ~ ~: \text{ elsewise}\hfill
		\end{gathered}  \right.
	\label{eq:Heff-lim0}
\end{equation}
is obtained. It can be seen, that the 
deterministic limit cannot be used at $\sigma=0$.
Surprisingly, small noise limit  leads to a vanishing
of current velocity in a finite interval around the fixed point $\tht_-$ and not just at this point.
A similar derivation is possible for first passage velocity, as for example seen
by mirror symmetry.

\begin{figure}[h]
   \centering
   \includegraphics[width=0.49\textwidth]{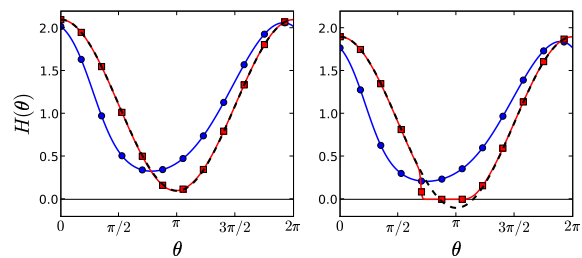}
   \caption{ Effective velocity at small $\sqg$ (red squares) converges to $h$ (dashed black line) for noise-perturbed oscillations (left plot),
   	but for noise-induced oscillations becomes discontinuous (right plot) as in formula \eq{Heff-lim0}. 
	Furthermore, the total area under the curve is independent of $\sigma$ (formula \eq{Int2}). Red squares: $\sqg=0.1$, Blue circles: $\sqg=0.7$. 
	Parameter $a$ is chosen to be $1.1$ (left plot), and $0.9$ (right plot).
  }
	\label{fig:1}
\end{figure}

In Fig.~\fig{1} the peculiar nature of $H(\tht)$ is illustrated for the theta
model. If noise is small, the effective velocity becomes discontinuous for
noise-induced oscillations where $|a|<1$ (right plot), whereas for
noise-perturbed oscillations at $|a|\ge1$, $H(\tht)$ converges to $h(\tht)$
for all $\tht$ (left plot). 

\subsection{Estimating frequency for large noise} \label{ssec:estifreq}

For large noise, an estimation of frequency can be troublesome when one has to
rely on Monte-Carlo simulations. Here, we want to provide a formula that allows
for an estimation of frequency at large noise amplitude $\sigma\to\infty$,
when given an effective velocity at arbitrary $\sigma$.

In the limit of large noise the probability density becomes uniform. By the
definition of, for example, current velocity \eq{flowvelocity} it is seen that:
\begin{equation}
	\lim_{\sigma\to\infty}I = \lim_{\sigma\to\infty}\frac{1}{2\pi}\int_0^{2\pi}H(\theta)~\dd\theta = \lim_{\sigma\to\infty}\omega = \w_\infty~.
	\label{eq:Int0}
\end{equation}
Now, it is shown that integral $I$ does not depend on $\sigma$.
Using equation \eq{Heff2}, current velocity is represented as
$H = h-\Gam\left[\ln P\right]'$ which yields for the integral
$\del{\Gam}I=0$. Therefore, $I$ does not depend on $\sigma$,
and we have
\begin{equation}
	\frac{1}{2\pi}\int_0^{2\pi}H(\theta)~\dd\theta = \w_\infty~,
	\label{eq:Int1}
\end{equation}
where $H$ may be given for arbitrary noise intensity $\sigma$.
With equation \eq{Heff-lim0} it can also be verified that
\begin{equation}
	\frac{1}{2\pi}\int_0^{2\pi}h(\theta)~\dd\theta = \w_\infty~.
	\label{eq:Int2}
\end{equation}
For additive noise it is seen that the area under the function $H(\tht)$ is a
preserved quantity and equal to $2\pi\w_\infty$. The same result can
be obtained for the first passage velocity, straight-forwardly.

\section{Stochastic phase resetting} \label{sec:stochPRC}

In the realm of phase resetting one is interested in the response of an
oscillator to a brief stimulus, a kick, applied to the system at certain
protophase $\alpha$ with certain strength and direction $\vv{k}$. The freely
rotating period $T$ is compared to the period $T_\vv{k}(\alpha)$ wherein the
kick is applied, by computing the \textit{phase resetting curve}~\cite{Canavier2006}
\begin{equation}
	\Delta\psi(\alpha,\vv{k}) = 2\pi\frac{T-T_\vv{k}(\alpha)}{T}.
	\label{eq:prc}
\end{equation}
It gives the shift in uniformly rotating phase the oscillator experiences due
to the kick. For example equation \eq{trafo1-2} provides the phase resetting
curve of current model \eq{effOsci}, whereas \eq{trafo2-2} provides it for
first passage model \eq{fpm} when for each setting $\tht_2=\tht_1+k$. Up to
here, quantities appearing in equation \eq{prc} are only well-defined for
deterministic systems that show limit-cycle oscillations. In this section we
extend the applicability of formula \eq{prc} to stochastic phase oscillators
(cf.~\cite{Ermentrout-Saunders-06}) and discuss the results in terms of
effective phase theory.

A kick applied at a time $t'$ with scalar strength $k$ representing a brief
stimulus is introduced to our stochastic phase oscillator \eq{generalOsci} by
formula
\begin{equation}
	\dot{\tht}(t)=h(\tht(t))+g(\tht(t))\xi(t)+k\delta(t'-t)~,
	\label{eq:Osciwithkick}
\end{equation}
Phase resetting is computed by a comparison of the kicked and unkicked
stochastic phase oscillator. For this, quantities appearing in formula \eq{prc}
are interpreted as follows: Quantity $T$ is given by mean period $T=2\pi/\w$
computed for the unkicked oscillator. Quantity
$T_k(\alpha)$ is computed for the kicked oscillator as the mean conditional first
passage time starting at $\tht=\alpha+k$ (value just after
kick), to reach boundary $\tht=\alpha+2\pi$. Here, the mean shall be taken with
respect to noise. 

By the above interpretation of formula \eq{prc} for kicked stochastic
oscillators, phase resetting can be calculated using mean first passage times.
From formula \eq{meanT} it is deduced that
$T(\alpha,\beta)+T(\beta,\delta)=T(\alpha,\delta)$. This is related to the
Markov property of $\tht(t)$, and it allows us to express mean conditional
first passage time as
\begin{equation}
	T_k(\alpha) = T-T(\alpha,\alpha+k)~,
	\label{eq:repT-kalpha}
\end{equation}
For equation \eq{prc}, it follows $\Delta\psi(\alpha,k)=2\pi T(\alpha,\alpha+k)/T$.
This can be rewritten in terms of $R$ using equation \eq{fpd} as
\begin{equation}
	\Delta\psi(\alpha,k) = 2\pi\intl_{\alpha}^{\alpha+k}R(\tht)~\dd\tht~.
	\label{eq:prcR}
\end{equation}
This is the exact formula of phase resetting curve $\Delta\psi(\alpha,k)$ for a
general stochastic phase oscillator \eq{generalOsci}.


It is seen that, phase resetting curve
\begin{equation}
	\Delta{\ph}(\alpha,k)=\lim_{\eps\to0}\left[\ph(t'+\eps)-\ph(t'-\eps)\right]_{\tht(t')=\alpha}=2\pi\intl_{\alpha}^{\alpha+k}P(\tht)~\dd\tht~,
	\label{eq:wrongPRC}
\end{equation}
derived for the current
model does not correspond to that of the stochastic phase oscillator, whereas
phase resetting curve 
\begin{equation}
	\Delta{\psi}(\alpha,k)=\lim_{\eps\to0}\left[\psi(t'+\eps)-\psi(t'-\eps)\right]_{\tht(t')=\alpha}=2\pi\intl_{\alpha}^{\alpha+k}P(\tht)~\dd\tht~,
	\label{eq:rightPRC}
\end{equation}
derived from first passage model \eq{fpm} yields the correct formula \eq{prcR}.
Let us explain why the current model fails. In section \ref{sec:currentModel},
it was seen that current velocity is given by $H(\tht)=J/P$, which leads to the
stationary solution \eq{flowvelocity}.  However, the stationary state is broken
in the phase resetting procedure, where the time evolution of $\tht(t)$ starts from the definite value
 $\tht=\alpha+k$.  Consider for example the moment, right after the
resetting, where $P(\tht)=\delta(\tht-\alpha-k)$. The probability flux is
calculated by integrating equation \eq{flux} and one obtains
$J(\alpha)=h(\alpha+k)$.  Deviations to the stationary solution \eq{flowvelocity}
are most prominent in the excitable regime where $J(\alpha+k)=h(\alpha+k)<0$. In
this case, the time-dependent current model 
\begin{equation}
	\dot{\tht} = \frac{J(\tht,t)}{P(\tht,t)}~,
	\label{eq:tdepH}
\end{equation}
has non-monotonic dynamics, and therefore does not yield a good phase
description.

\begin{figure}[h]
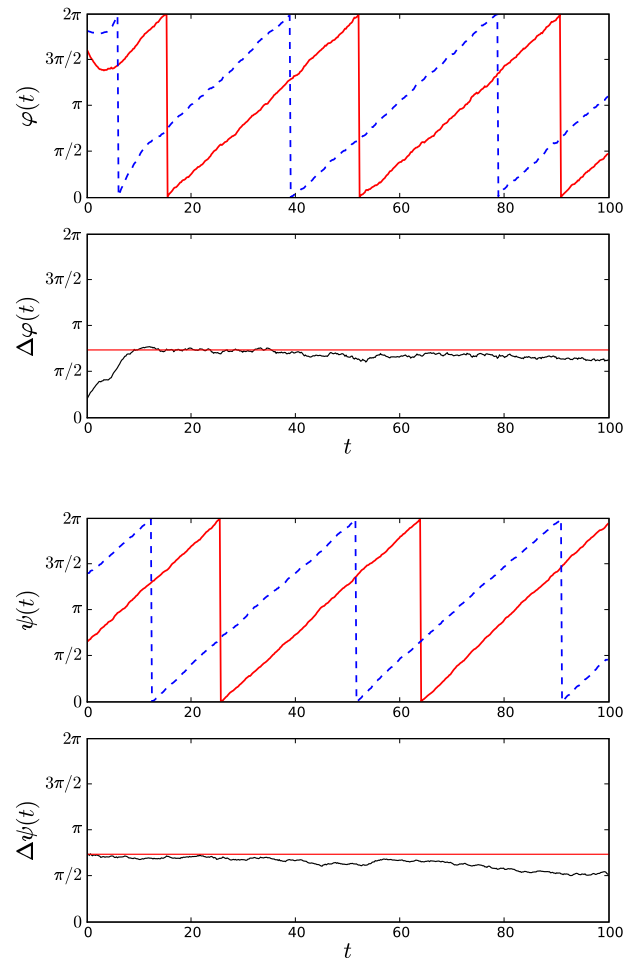

   \centering
   \includegraphics[width=0.45\textwidth]{sec5-flowBreakdown.epsi}\hfill\\~\\~\\
   \includegraphics[width=0.45\textwidth]{sec5-prc.epsi}
   \caption{Dynamics of $\ph(t)=S(\tht(t))$ and $\psi(t)=Z(\tht(t))$, averaged  over $1000$ representations starting
   		at two distinct initial conditions $\tht_{1/2}=\tht_+\pm0.2$ (top plots, $+$: blue dashed, $-$: red solid). Left column confirms that
		\eq{uniformPhase} does not hold for the phase resetting. For
		fixed initial conditions the initial phase difference $\Delta\ph(0)=S(\tht_2)-S(\tht_1)$
		widens (bottom plot, top Fig.) to the value $\Delta\psi=Z(\tht_2)-Z(\tht_1)$ (red horizontal line);
		for larger times phase difference diffuses away due to insufficient averaging.
   Right column confirms that
		equation \eq{rightPRC} gives the correct phase resetting curve for stochastic oscillators. The initial phase difference $\Delta\psi(0)=Z(\tht_2)-Z(\tht_1)$
		persists (bottom plot, bottom Fig.) at its theoretically predicted value. 
  }
	\label{fig:sec5-flowBreakdown}
\end{figure}

The failure of current model to predict the correct phase resetting curve is
illustrated in figure \fig{sec5-flowBreakdown} by noise-induced oscillations of
the theta model. For this, $1000$ representations starting from
$\tht_1=\tht_+-0.2$ (red solid line) and $\tht_2=\tht_++0.2$ (blue dashed line)
were calculated and transformed to uniformly distributed phase $\ph=S(\tht)$.
As before, $\tht_+$ is the unstable fixed point of corresponding arbitrary model.
For each group an average over representations was performed to obtain average
dynamics $\ph(t)$. Equation \eq{uniformPhase} predicts $\ph$ to be
rotating uniformly. Numerically, it is found however  that while
asymptotically uniformly rotating, initially $\ph(t)$ has systematic
non-uniformities such that initial phase difference
$\Delta\ph(0)=S(\tht_2)-S(\tht_1)$ widens. Asymptotically, it
reaches the value $\Delta\psi=Z(\tht_2)-Z(\tht_1)$ as correctly predicted by
stochastic phase resetting \eq{prcR} and first passage model \eq{rightPRC}. For
the stochastic phase oscillator with initial condition $\tht_1$ with
$h(\tht_1)<0$, it is furthermore seen that $\ddd{t}\langle\ph\rangle<0$
initially, as explained above. 
Performing the same procedure using transformation $\psi=Z(\tht)$ it is
confirmed that formula \eq{rightPRC} predicts the correct phase resetting curve
(see right plot).

\section{Conclusions}

Two useful formulations of effective phase theory are presented, each relying
on different definitions of phase velocity. While the concept of time spent on
average in an interval leads to current velocity, a definition based on first
passage times leads to first passage velocity. For noise-perturbed
oscillations, the two velocities converge for vanishing noise, whereas for
noise-induced oscillations they don't. Whereas the current model was shown to
be useful for a characterization of continuous coupling~\cite{Pikovsky2010} of
stochastic phase oscillators, it was seen in this article, that first passage
model gives a correct description of phase resetting. 

A natural extension of effective phase theory would be to construct models of
two-dimensional oscillators for which a phase variable and a corresponding
radius variable may be found. The extension is not straight-forward: A
deterministic current-type model for higher-dimensional oscillators must have a
meaning different from the one-dimensional case, because the correspondence
between distribution and probability density (condition (ii)) can not be drawn.
Here, also other interesting difficulties arise as described
in~\cite{Guralnik2008}.  On the other hand, it might be possible to construct a
first passage-type model exhibiting the correct phase resetting curve. In this
sense the approach already implies, that a perturbative approach, as applied
in~\cite{Yoshimura2008}, for a construction of a phase resetting curve is
expected to be biased. 

Together with these issues, it remains unresolved how to construct a
theoretical description of a stochastic phase oscillator under the influence of
an external force consisting of both, a continuous and a pulse-like
contribution. Furthermore, we expect the presented work to be useful for a
characterization of finite ensembles of oscillators with either continuous or
pulse-like coupling.


\textit{Acknowledgement} The work was supported by DFG via SFB 555 "Complex Nonlinear Processes".


\begin{thebibliography}{16}

\bibitem{Kuramoto-84}
Y.~Kuramoto, \emph{Chemical Oscillations, Waves and Turbulence} (Springer,
  Berlin, 1984)

\bibitem{Pikovsky-Rosenblum-Kurths-01}
A.~Pikovsky, M.~Rosenblum, J.~Kurths, \emph{Synchronization. A Universal
  Concept in Nonlinear Sciences.} (Cambridge University Press, Cambridge, 2001)

\bibitem{Kralemann-08}
B.~Kralemann, L.~Cimponeriu, M.~Rosenblum, A.~Pikovsky, R.~Mrowka, Phys. Rev. E
  \textbf{77}(6), 066205 (2008)

\bibitem{Canavier2007}
S.A. C.~C.~Canavier, Scholarpedia \textbf{2}(4), 1331 (2007)

\bibitem{Pikovsky2010}
J.T.C. Schwabedal, A.~Pikovsky, Phys. Rev. E \textbf{81}(4), 046218 (2010)

\bibitem{Nelson-66}
E.~Nelson, Phys. Rev. \textbf{150}(4), 1079 (1966)

\bibitem{Ermentrout2008}
G.B. Ermentrout, Scholarpedia \textbf{3}(3), 1398 (2008)

\bibitem{Risken1989}
H.Z. Risken, \emph{The Fokker--Planck Equation} (Springer, Berlin, 1989)

\bibitem{Reimann2001}
P.~Reimann, C.~Van~den Broeck, H.~Linke, P.~Hanggi, J.M. Rubi, A.~Perez-Madrid,
  Phys. Rev. Lett. \textbf{87}(1), 010602 (2001)

\bibitem{Pikovsky-Kurths-97}
A.S. Pikovsky, J.~Kurths, Phys. Rev. Lett. \textbf{78}(5), 775 (1997)

\bibitem{Goldobin2005}
D.S. Goldobin, A.~Pikovsky, Phys. Rev. E \textbf{71}(4), 045201 (2005)

\bibitem{Just2003}
W.~Just, H.~Kantz, M.~Ragwitz, F.~Schm\"user, Europhys. Lett. \textbf{62}(1),
  28 (2003)

\bibitem{Canavier2006}
C.C. Canavier, Scholarpedia \textbf{1}(12), 1332 (2006)

\bibitem{Ermentrout-Saunders-06}
B.~Ermentrout, D.~Saunders, J. Comput. Neuroscience \textbf{20}(2), 179 (2006)

\bibitem{Guralnik2008}
Z.~Guralnik, Chaos \textbf{18}, 033114 (2008)

\bibitem{Yoshimura2008}
K.~Yoshimura, K.~Arai, Phys. Rev. Lett. \textbf{101}, 154101 (2008)

\end{thebibliography}

\end{document}